\documentclass[preprint,times]{elsarticle}
 
\usepackage{lineno,hyperref} 
\usepackage{ulem}

\providecommand{\adsurl}[1]{\href{#1}{ADS}}

\usepackage[figure,figure*]{hypcap} 
\modulolinenumbers[5] 
\journal{New Astronomy Reviews}
\usepackage{color}

\def\ltsima{$\; \buildrel < \over \sim \;$}
\def\lsim{\lower.5ex\hbox{\ltsima}}
\def\gtsima{$\; \buildrel > \over \sim \;$}
\def\gsim{\lower.5ex\hbox{\gtsima}}

\begin{document}

\begin{frontmatter}

\title{Kepler-78 and the Ultra-Short-Period Planets}

\author[princeton]{Joshua N.\ Winn}
\ead{jnwinn@princeton.edu}

\author[roberto]{Roberto Sanchis-Ojeda}
\ead{rsanchis86@gmail.com}

\author[mit]{Saul Rappaport}
\ead{sar@space.mit.edu}

\address[princeton]{Princeton University, Princeton, NJ 08540, USA}
\address[roberto]{Netflix, Los Gatos, CA 95032, USA}
\address[mit]{Massachusetts Institute of Technology, Cambridge, MA 02139, USA}

\begin{abstract}

Compared to the Earth, the exoplanet Kepler-78b has a similar size
(1.2\,$R_\oplus$) and an orbital period a thousand times shorter (8.5
hours). It is currently the smallest planet for which the mass,
radius, and dayside brightness have all been measured.  Kepler-78b is
an exemplar of the ultra-short-period (USP) planets, a category
defined by the simple criterion $P_{\rm orb} < 1$~day.  We describe
our Fourier-based search of the {\it Kepler} data that led to the
discovery of Kepler-78b, and review what has since been learned about
the population of USP planets.  They are about as common as hot
Jupiters, and they are almost always smaller than 2\,$R_\oplus$. They
are often members of compact multi-planet systems, although they tend
to have relatively large period ratios and mutual inclinations. They
might be the exposed rocky cores of ``gas dwarfs,'' the planets
between 2--4\,$R_\oplus$ in size that are commonly found in somewhat
wider orbits.
  
\end{abstract}

\begin{keyword}
\texttt{planets \sep time-series photometry}
\end{keyword}

\end{frontmatter}


\section{Introduction}

\label{sec:introduction}

\begin{figure*}[h!]
 \begin{center}
 \leavevmode
\includegraphics[keepaspectratio=true, width=5.0in]{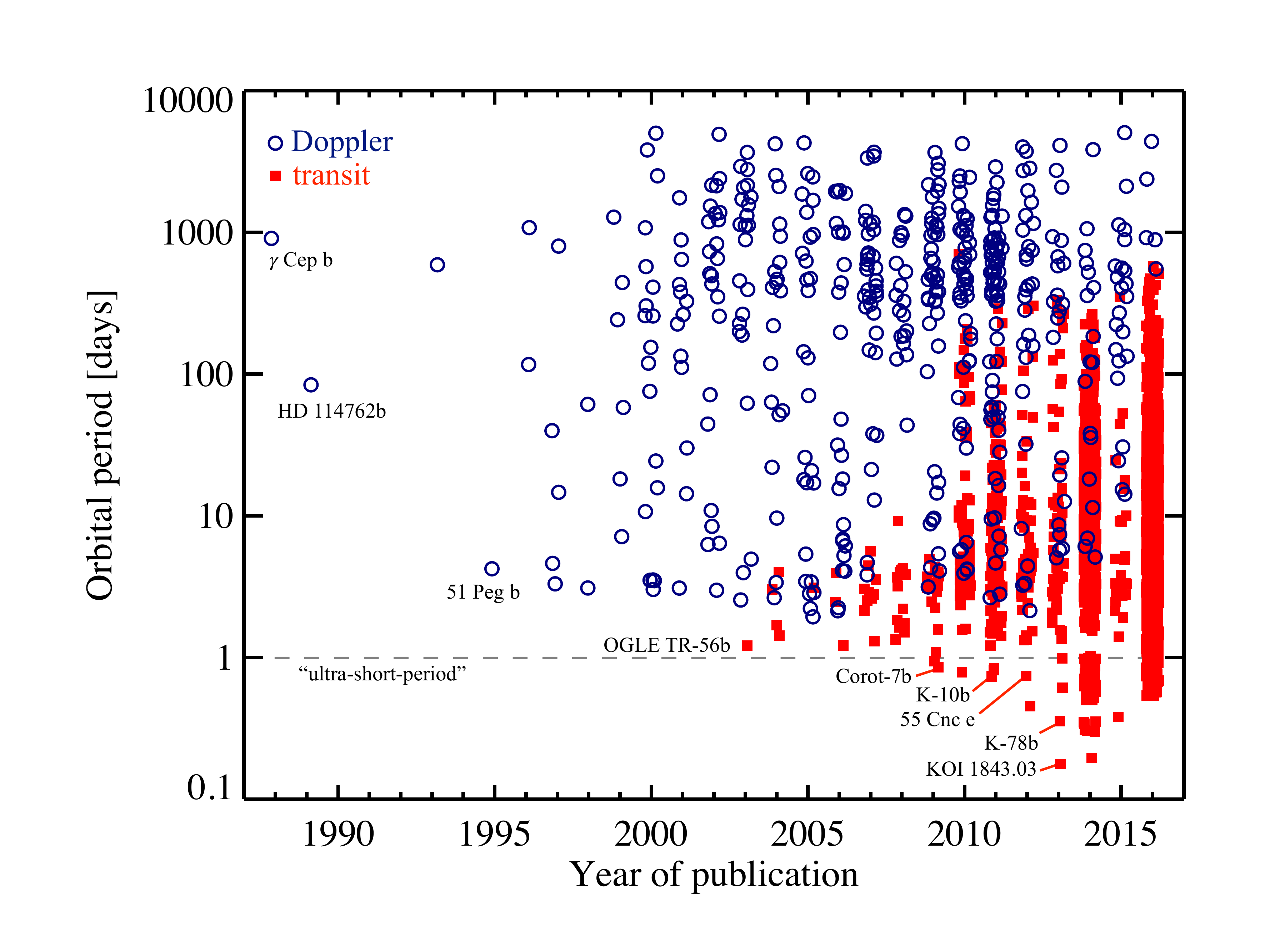}
 \end{center}
 \vspace{-0.4in}
 \caption{ Orbital periods of planets discovered through the Doppler
   and transit methods, versus the year of publication. Data were
   obtained from {\tt exoplanets.org} and other catalogs in the
   literature \cite{Jackson+2013,SanchisOjeda+2014}.  Small random
   shifts have been applied to the horizontal positions to allow the
   data points to be more clearly distinguished.
  \label{fig:period_versus_time}}
\end{figure*}

One of the earliest surprises of exoplanetary science was the
existence of planets with very short orbital periods.  The main
pre-exoplanet theory of planet formation predicted that gas giant
planets like Jupiter could only form in wide orbits, more than 2--3
times the size of Earth's orbit around the Sun.  Then in 1995, the
star 51~Pegasi was found to host a Jupiter-mass planet with an orbit
only 5\% of the size of Earth's orbit \citep{MayorQueloz1995}. The
planet rushes around the star every 3.5 days, an orbital period far
shorter than Jupiter's 4300-day period, or even Mercury's 88-day
period.  How planets attain such short periods is the oldest
unresolved problem in the field.

After 51~Peg, a good place to continue the story is in 2003 with the
discovery of OGLE-TR-56b \cite{Konacki+2003} (see
Figure~\ref{fig:period_versus_time}).  The initial report of this
planet was met with skepticism because of the unusually small orbital
distance of 0.023~AU and short orbital period of 1.2 days.
Practitioners of the Doppler technique wondered why the very first
planet to emerge from transit surveys would have such a short period
while the longstanding Doppler surveys had not yet found any such
planets.

The resolution of this problem was that transit surveys are even more
strongly biased toward short periods than Doppler surveys
\citep{Gaudi+2005}. Consider a transit survey in which all the nearby
stars within some field of view are repeatedly imaged, allowing the
flux $F$ of each star to be measured with a fractional uncertainty
proportional to $F^{-1/2}$.  If all the transit signals exceeding a
certain signal-to-noise threshold can be detected, then the number of
stars for which a planet of period $P$ would produce a detectable
transit signal scales as $P^{-5/3}$ \cite{Pepper+2003,Gaudi+2005}.  A
survey designed to search a certain number of stars for Earth-sized
planets with a period of 365~days --- such as the {\it Kepler} survey
--- is capable in principle of searching $(365)^{5/3} \approx
19{,}000$ times as many stars for Earth-sized planets with a period of
1~day.  Hence, it is possible to find very short-period planets even if
they are exceedingly rare.

A transit survey with the {\it Hubble Space Telescope} in 2004 led to
the detection of five giant-planet candidates with periods shorter
than one day, which the authors referred to as ``ultra-short-period
planets'' \cite{Sahu+2006}.  However, it was difficult to have much
confidence that these candidates were truly planets, because of the
limited amount of data collected (7 days) and the faintness of the
host stars ($V=22$--26~mag).  As ground-based transit surveys made further
progress, it became clear that giant planets with such short periods
are very rare.  By early 2018, the ground-based transit surveys had
discovered 68 giant planets with periods between 2 and 3 days, but
only 37 with periods between 1 and 2 days, and a mere six with periods
shorter than one day.  This is despite the strong selection bias
favoring the shortest periods.

The next major advance came in 2009 during the European {\it Corot}
mission, the first spaceborne transit survey.  That year saw the
announcement of Corot-7b \cite{Leger+2009}, which was then the
smallest known transiting planet (1.7\,$R_\oplus$) and had the
shortest known orbital period (0.85~days).  A couple of years later,
the NASA {\it Kepler} mission found a similar planet, Kepler-10b, with
a radius of 1.4\,$R_\oplus$ and an orbital period of 0.84~days
\cite{Batalha+2011}.  In between these discoveries was a curious
episode involving the innermost planet of the 55~Cnc system.  The
planet was initially discovered through the Doppler technique
\cite{McArthur+2004,Fischer+2008}, but the period was misidentified as
2.8 days due to aliasing.  Subsequent analysis of the Doppler data
\citep{DawsonFabrycky2010} and the detection of transits with
space-based photometry \cite{Winn+2011,Demory+2011} showed that the
true period is 0.74~days.

For Sun-like stars, it is now clear that planets with periods shorter
than one day occur just as frequently as hot Jupiters with periods
ranging from 1 to 10 days.  The reason that the ultra-short-period
(USP) planets had previously escaped detection is that they are small,
and produce signals that are difficult to detect without the precision
of space telescopes.  The USP planets have also been called ``hot
Earths,'' or more evocatively, ``lava worlds'' \cite{Leger+2011}, as
their dayside surface temperatures are higher than the melting point
of most rock-forming minerals.  The current record holder is
KOI-1843.03, which circles its star every 4.2 hours
\cite{OfirDreizler2013,Rappaport+2013}.\footnote{Even though no {\it
    Kepler} number has been assigned, KOI-1843.03 is likely to be
  planet.  The signal was validated as a probable planet through the
  usual tests \cite{Rappaport+2013}.  In addition, any sources with
  multiple detected transit signals, such as KOI-1843, are likely to
  represent genuine planetary systems \cite{Lissauer+2012}.} Not far
behind is K2-137b, with an orbital period of 4.3 hours
\cite{Smith+2018}.

When trying to understand a process as complex as planet formation,
the most extreme cases are often the most revealing.  This is one
reason why the study of USP planets is rewarding.  They may help us to
understand the formation and orbital evolution of short-period
planets, as well as star-planet interactions, atmospheric erosion, and
other phenomena arising from strong irradiation and strong tidal
forces. In addition there are practical advantages to studying USP
planets. They are easier to detect than planets of the same size in
wider orbits. Their masses and sizes, the most basic inputs to
theories of planetary interiors, are easier to measure.  They are
sometimes hot enough to emit a detectable glow, enabling observations
to determine their surface temperature and reflectivity, which is
usually impossible for wider-orbiting planets.

The defining criterion of $P_{\rm orb} < 1$~day is arbitrary.  We
chose it because 1 is a nice round number, and because planets with
such short periods were relatively unexplored at the time of our
survey.  Nature does not seem to produce any sharp astrophysical
distinction between planets just inside or outside of the one-day
boundary.  A more meaningful boundary might be at about 10 days,
beyond which we start to see differences in the planetary occurrence
rate and the mean metallicity of the host stars.

In the spirit of this special issue, Section~\ref{sec:search} tells
the story of Kepler-78b, a planet with Earth-like proportions and an
orbital period of 8.5 hours. This planet is important because it is
one of the very smallest planets for which both the mass and radius
have been measured to better than 20\%.  It is also the smallest
planet for which the brightness of the dayside has been determined,
through the detection of planetary occultations.  Kepler-78b was one
of the first USP planets that emerged from our systematic search
through the {\it Kepler} data.  We also take this opportunity, in
Section~\ref{sec:observations}, to review the searches undertaken by
other groups, and the growing collection of related investigations
into this enigmatic population of planets.  Section~\ref{sec:other}
describes some other intriguing ultra-short-period phenomena that
might, or might not, be related to planets.

\section{Kepler-78}
\label{sec:search}

The first appearance in the literature of the star now known as
Kepler-78 is in the second data release of the {\it Kepler} Eclipsing
Binary catalog \cite{Slawson+2011}, published on October 12, 2011.  It
was listed with {\it Kepler} Input Catalog (KIC) number 8435766, and
deemed a detached eclipsing binary with a period of 0.71~days (17
hours).  We do not know what led to this classification, or why the
period was found to be twice as long as the true value.  Occasional
errors of this type are to be expected in any catalog based on
automated analysis of a large database.  In any case, we were
unaware of this entry in the eclipsing binary catalog.  The star came
to our attention through a different route.

We had been searching the {\it Kepler} data for transiting planets
based on the Fourier Transform (FT) of the light curve, instead of the
more widely used Box-fitting Least Squares (BLS) spectrum
\cite{Kovacs+2002}.  Our motivation was to perform a complementary
search for planets in a regime where the standard {\it Kepler}
pipeline was having difficulty at that time.  The BLS method was
invented specifically for the detection of transit signals: brief
drops in brightness modeled as a ``box'' (rectangular-pulse) function
with duration $T$ and period $P$.  The motivation for the FT method
might not be obvious, because the FT of a box function has power
spread over many harmonics in addition to the fundamental frequency.
Specifically, it consists of a series of peaks with $f_n = n/P$,
modulated by the amplitude function
\begin{equation}
A(f) = \frac{T\sin (\pi f T)}{\pi f T}
\end{equation}
Most of the power is concentrated in the frequency
range from zero to the first null at $1/T$, within which the number
of harmonics is approximately $P/T$. For a transiting
planet around a Sun-like star,
\begin{equation}
\frac{P}{T} \sim 500~\left( \frac{P}{1~{\rm year}} \right)^{2/3}.
\end{equation}
Splitting the signal among hundreds of harmonics sound like a terrible
idea.  The peaks would be easily lost in the noise.  But for a planet
with an 8-hour period, there are only 5--6 strong harmonics, making the
FT method reasonably effective. We also found it to
have some practical advantages: it is fast and easy to compute, and
it performs well in the presence of the most common types of stellar
variability.

By early March of 2013, one of us (S.A.R.) had inspected the FTs of
about 10{,}000 {\it Kepler} light curves.  Planet candidates were
identified based on the presence of 5 or 6 strong harmonics.  In many
cases, periodic fading events turn out to be caused by eclipsing
binary stars, rather than transiting planets. Sometimes these ``false
positives'' could be recognized from the presence of
subharmonics, produced by alternating eclipse depths.  The resulting
list of 93 candidates with strong harmonics and no subharmonics were
then vetted in the usual ways \cite{Jenkins+2010,Morton+2011}.  We
produced a phase-folded light curve after filtering out any stellar
variability, and confirmed that there was no detectable alternation in
eclipse depths, no detectable ellipsoidal light variations, and no
detectable motion of the center of light in the {\it Kepler} images
that would have suggested a blend between an eclipsing binary and a
brighter foreground star.

We realized on March 7 that KIC\,8435766 was special.  It earned one
of the only two ``A+'' grades that were awarded during the visual
inspection\footnote{The other A+ grade went to KOI-1843.03, which we
  later learned had already been reported in the
  literature \cite{OfirDreizler2013}.}; it had the brightest host star
($m_{\rm Kep}= 11.5~{\rm mag})$; and it showed evidence for tiny dips in between
eclipses that were consistent with planetary occultations.  It was an
unexpected gift.  We thought the brightest {\it Kepler} stars had
already been thoroughly picked over by other groups.  After a few days
of further analysis, we contacted David Latham to request
spectroscopy, and by the beginning of April we could rule out
radial-velocity variations at the 100~m\,s$^{-1}$ level, placing the
mass of the transiting companion within the planetary regime.  Our
paper was submitted on May 15, 2013.  Soon after, on July 12, the
system was designated Kepler-78 by the staff at the NASA Exoplanet
Archive.\footnote{Although it appears in the published paper as
  ``Kepler-XX'' because we forgot to inform the {\it ApJ} editorial
  staff of the name change before it was too late.}

The spectra obtained by Latham's group confirmed that the host star
was a late G dwarf and suggested it might be suitable for precise
Doppler monitoring.  The possibility beckoned of measuring the mass of
a nearly Earth-sized planet.  On May 20 and 21, a conference was held
at the Harvard-Smithsonian Center for Astrophysics in honor of
Latham's distinguished career.  During a lunch break we met with
members of the California Planet Search (CPS) to discuss the
possibility of using the Keck~I 10\,m telescope and the HIRES
spectrograph to measure the mass of the planet.  During the same
meeting, unbeknownst to us, the members of the HARPS-North consortium
were also planning to conduct precise Doppler observations at the
nearest possible opportunity.  Given the brightness of the star and
the high level of confidence in the planetary nature of Kepler-78b, it
was not too surprising that two different teams embarked on campaigns
to measure the mass of the planet.

Both teams began collecting data.  It soon became clear that the
biggest hurdle in measuring the mass of Kepler-78b would be the
starspot activity of the host star.  The {\it Kepler} data showed
fluctuations in total light of 1\%, presumably due to starspots
rotating across the star's visible hemisphere.  This level of activity
leads to spurious Doppler shifts on the order of 10~m\,s$^{-1}$, much
larger than the expected planetary signal of 1--2~m\,s$^{-1}$.
Fortunately, the rotation period (12.5 days) is much longer than the
orbital period (0.36~day), allowing a clear separation of timescales.
It proved possible to detect the planet-induced radial-velocity signal
through intensive observations on individual nights, during which the
effects of stellar rotation are minimal.

By early June, the CPS and HARPS-N teams had learned of each other's
plans.  An arrangement was made to submit our journal articles at the
same time, but without sharing any results until just beforehand.  The
two independent results for the planet mass agreed to within
1-$\sigma$, despite the different approaches that were taken to cope
with the effects of stellar activity.  The articles were submitted on
September 25, 2013 \cite{Howard+2013,Pepe+2013}.  Since then, other
groups have confirmed the robustness of these measurements by
combining both datasets and using different analysis techniques
\cite{Hatzes2014,Grunblatt+2015}.  The most recent such study found
$R_p=1.20\pm 0.09$\,$R_\oplus$ and $M_p=1.87\pm 0.27$\,$M_\oplus$,
giving a mean density of $6.0^{+1.9}_{-1.4}$\,g\,cm$^{-3}$
\cite{Grunblatt+2015}. This is consistent with the Earth's mean
density of 5.5\,g\,cm$^{-3}$, although the uncertainty is large enough
to allow for a wide range of possible combinations.  Using a simple
model consisting only of rock and iron, the iron fraction is
constrained to be $0.32\pm 0.16$.

\begin{figure*}[h!]
 \begin{center}
 \leavevmode
\includegraphics[keepaspectratio=true, width=4.5in]{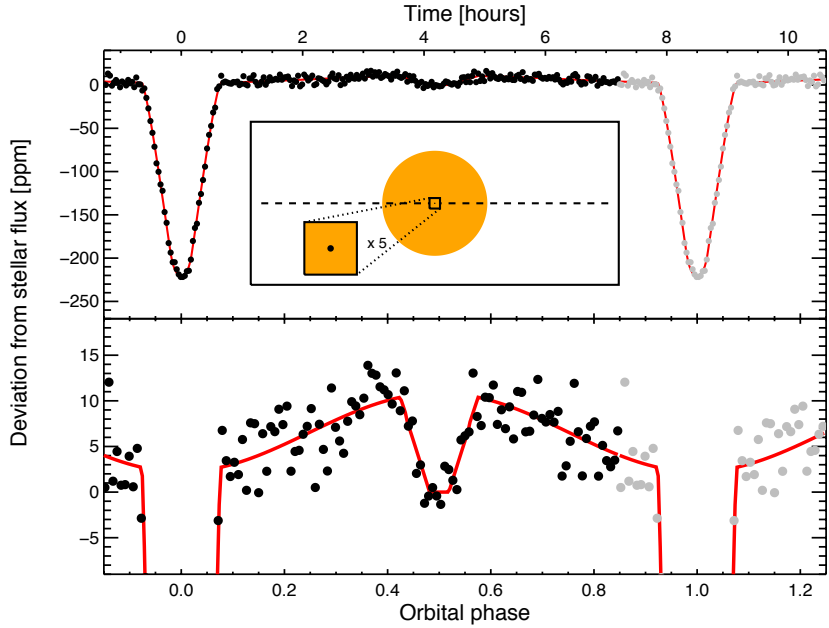}
 \end{center}
 \vspace{-0.15in}
 \caption{Transit light curve of Kepler-78b \cite{SanchisOjeda+2013}.
   The upper panel shows the phase-folded light curve based on nearly
   4 years of {\it Kepler} observations (black dots).  The sub-panel
   shows a schematic of the planetary system: the dashed line
   represents the projected orbit of the planet, and the planet itself
   is barely visible in the middle of the star. The lower panel gives
   a clearer look at the occultation signal and the sinusoidal
   variation produced by some combination of reflected and reprocessed
   light. \label{fig:kepler78_foldedlc}}
\end{figure*}

The brightness of the host star has enabled other interesting
measurements.  Any long-term change in orbital period due to tidal
effects, planet-planet interactions, or other reasons must satisfy
$|P/\dot{P}| > 2.8$~Myr.  The detection of occultations and the
planet's ``phase curve'' (the gradual light variations that are seen
with the same period as the planetary orbit) help to limit the
possibilities for the planet's visual albedo and surface temperature.
If the Bond albedo is 0.5, the dayside temperature must be
$\approx$\,2600\,K, while if the albedo is very small then the
temperature could be as high as 3000\,K.  The nightside temperature is
less than $2700\,{\rm K}$ (3-$\sigma$).  Future observations at
multiple wavelengths, perhaps with the {\it James Webb Space
  Telescope}, will lead to better constraints.

The relatively young age of the star (750\,Myr) and high level of
activity have allowed some investigators to characterize the stellar
magnetic field.  Spectropolarimetry was used to infer a surface
magnetic field strength of 16\,G for the host star, by exploiting the
Zeeman effect \citep{Moutou+2016}.  It has been suggested that
planet-star interactions could lead to a detectable modulation in
magnetic activity through a mechanism similar to the electrodynamic
coupling between Jupiter and Io \cite{Laine+2012}. No such modulation
has yet been detected.

\section{The USP planet population}
\label{sec:observations}

\subsection{Physical characteristics}
\label{subsec:basic}

Having indulged ourselves in telling the story of Kepler-78, we turn
to the more scientifically relevant task of summarizing what has been
learned about USP planets in general.  To get oriented, we begin with
some basic physical considerations.  We also allude to some of the
more sophisticated models that have been developed for extremely hot
rocky planets, in response to the discoveries of Corot-7b, Kepler-10b,
55~Cnc~e, and Kepler-78b.

To set the scale of the orbit, we apply Kepler's third law,
\begin{equation}
  a = 0.0196\,{\rm AU}~\left( \frac{P_{\rm orb}}{1~{\rm day}} \right)^{2/3}
  \left( \frac{M_\star}{M_\odot} \right)^{1/3},~~
  \frac{a}{R_\star} = 4.2~\left( \frac{P_{\rm orb}}{1\,{\rm day}} \right)^{2/3}
  \left( \frac{\rho_\star}{\rho_\odot} \right)^{1/3}.
\end{equation}
For these fiducial parameters, the angular diameter of the star in the
sky of the planet is $27^\circ$, i.e., fifty times wider than the Sun in
our sky.

At this short range, tidal interactions lead to relatively rapid
orbital and spin evolution.  In the constant-lag-angle model of the
equilibrium tide, the timescale for orbital circularization is
\cite{GoldreichSoter1966,Patra+2017}
\begin{equation}
  \frac{e}{\dot{e}} \sim 1.7\,{\rm Myr}
  \left(\frac{Q_{\rm p}'}{10^3}\right)
  \left(\frac{M_{\rm p}/M_\star}{M_\oplus/M_\odot}\right)
  \left(\frac{R_\star/R_{\rm p}}{R_\odot/R_\oplus}\right)^5
  \left( \frac{\rho_\star}{\rho_\odot} \right)^{5/3}
  \left( \frac{P_{\rm orb}}{1~{\rm day}} \right)^{13/3},
\end{equation}
where $Q_{\rm p}'$ is the modified tidal quality
factor characterizing the dissipation rate of tidal oscillations,
scaled to a customary value for the Earth.
The timescale for the planet to achieve spin-orbit synchrony is even
shorter, by a factor on the order of $(R_{\rm p}/a)^2$.  Therefore
when we see a USP planet around a mature main-sequence star it
reasonable to assume (until proven otherwise) that the planet has a
circular orbit and a permanent dayside and nightside.  However, the
orbit of a terrestrial-mass USP planet does not have enough
angular momentum to spin up the star and achieve a stable
double-synchronous state.  Instead, the planet spirals into the star on
a timescale \cite{GoldreichSoter1966, Patra+2017}
\begin{equation}
  \frac{P}{\dot{P}} \sim 30\,{\rm Gyr}
  \left(\frac{Q_\star'}{10^6}\right)
  \left(\frac{M_\star/M_{\rm p}}{M_\odot/M_\star}\right)
  \left( \frac{P_{\rm orb}}{1~{\rm day}} \right)^{13/3}
  \left( \frac{\rho_\star}{\rho_\odot} \right)^{5/3},
\end{equation}
where $Q_\star'$ is the star's modified tidal quality parameter.
Since $Q_\star'$ is uncertain by at least an order of magnitude, it is
not clear whether tidal orbital decay is important on
astrophysical timescales.

A one-day planet around a Sun-like star intercepts a flux of
3.5\,MW\,m$^{-2}$, about 2600 times the flux of the Sun impinging on the
Earth. Under the simplifying assumption that all the incident energy
is re-radiated locally, the planet's surface at the substellar point
(high noon) is raised to a temperature of 2800\,K.  This is not far
from the temperature of the glowing tungsten filament in an
incandescent light \cite{Leger+2011}. It is also hot enough to melt
silicates and iron, a fact which has led to theoretical work on the
properties of the resulting lava oceans
\cite{Leger+2011,Rouan+2011,Kite+2016} and mantle convection
\cite{Gelman+2011,Wagner+2012}.

During the first $10^7$ years of a star's active youth, a USP planet
would be bathed in ultraviolet and X-ray radiation.  The gas near the
XUV photosphere would be heated to such a degree that the pressure
gradient would drive a hydrodynamic wind, leading to atmospheric
escape \cite{LopezFortney2016, OwensWu2016, ChenRogers2016}.  This
would lead to a complete loss of any hydrogen-helium
atmosphere. Unless the planet has an envelope of water vapor or other
elements heavy enough to be retained \cite{Lopez+2012}, the bare solid
surface would sit beneath a very thin atmosphere with a maximum
pressure of order $10^{-5}$~atm \cite{Leger+2011}.  Models which track
the chemistry of Earth's crust as it is heated to a temperature of
1500--3000\,K suggest that the outgassed atmosphere would be mainly
composed of Na, O$_2$, O, and SiO \cite{SchaeferFegley2009}.

\subsection{Detections}

Several groups have undertaken systematic searches of the {\it Kepler}
data for short-period planets, only a few of which were specifically
designed for periods shorter than about one day.  In 2014, we and our
collaborators published a list of 106 USP planet candidates based on the
concatenation of our own detections as well as those of other groups
\cite{OfirDreizler2013,Jackson+2013,Huang2013a}.  The characteristics
of the stars and planets were later clarified based on high-resolution
optical spectroscopy, and a few false positives were uncovered
\cite{Winn+2017}.

In 2016, the population of USP planets was noted by other
investigators.  One group confirmed that the sample of strongly irradiated planets does not
include many with sizes exceeding 2~$R_\oplus$ \cite{Lundkvist+2016}.
Their sample consisted of {\it Kepler} systems for which we have
unusually good knowledge of the stellar and planetary sizes, thanks to
the detection of asteroseismic oscillations.  Another study concluded
that at least 17\% of the ``hot Earths'' detected by {\it Kepler} have
a different radius/period distribution than the planets in the
collection of {\it Kepler} multi-planet systems
\cite{SteffenCoughlin2016}.

In addition, over the last few years, a fresh sample of USP planets
has been found using data from the ongoing NASA {\it K2} mission.  A
systematic search was undertaken by the Short Period Planets Group
\cite{Adams+2016}, resulting in 19 candidates.  Other groups have
validated and characterized about a dozen additional candidates
\cite{Dressing+2017,Barragan+2018,Guenther+2017,Dai+2017,Malavolta+2018,Smith+2018}.

\subsection{Occurrence rate}

About one out of 200 Sun-like stars (G dwarfs) has an
ultra-short-period planet.  This result is based on a systematic and
largely automated search of the {\it Kepler} data using our FT
pipeline, calibrated with inject-and-recover simulations
\cite{SanchisOjeda+2014}.  The simulations showed that the efficiency
of detecting planets larger than $2\,R_\oplus$ was higher than $90\%$,
and that it dropped below $50\%$ for planets smaller than
$1\,R_\oplus$.  After correcting for this sensitivity function, and
for the transit probabilities, the occurrence rate was found to be
$(0.51\pm 0.07)\%$ for planets larger than 0.84\,$R_\oplus$ and
periods shorter than one day.

Among the other results of this survey was evidence for a strong
dependence of the occurrence rate upon the mass of the host star.  The
measured occurrence rate falls from $(1.1\pm 0.4)\%$ for M dwarfs to
$(0.15\pm 0.05)\%$ for F dwarfs.  There are still substantial
uncertainties in the occurrence rates for stars at either end of this
range, due to the relatively small number of detections.  It is
perhaps telling, though, that most of the USP planets found with {\it
  K2} data are around K and M dwarfs.

\begin{figure*}[h!]
 \begin{center}
 \leavevmode
\includegraphics[keepaspectratio=true, width=4.5in]{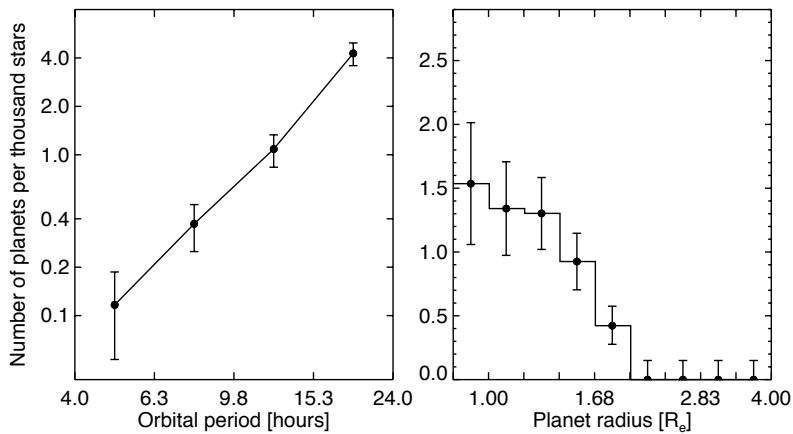}
 \end{center}
 \vspace{-0.7in}
 \caption{ Occurrence rates as a function of period and radius for USP
   planets orbiting G and K dwarfs \cite{SanchisOjeda+2014}. The
   period distribution is consistent with a power law.  The radius
   distribution shows a sharp decline at around
   $2\,R_\oplus$.\label{fig:pr}}
\end{figure*}

\begin{figure*}[h!]
 \begin{center}
 \leavevmode
\includegraphics[keepaspectratio=true, width=4.5in]{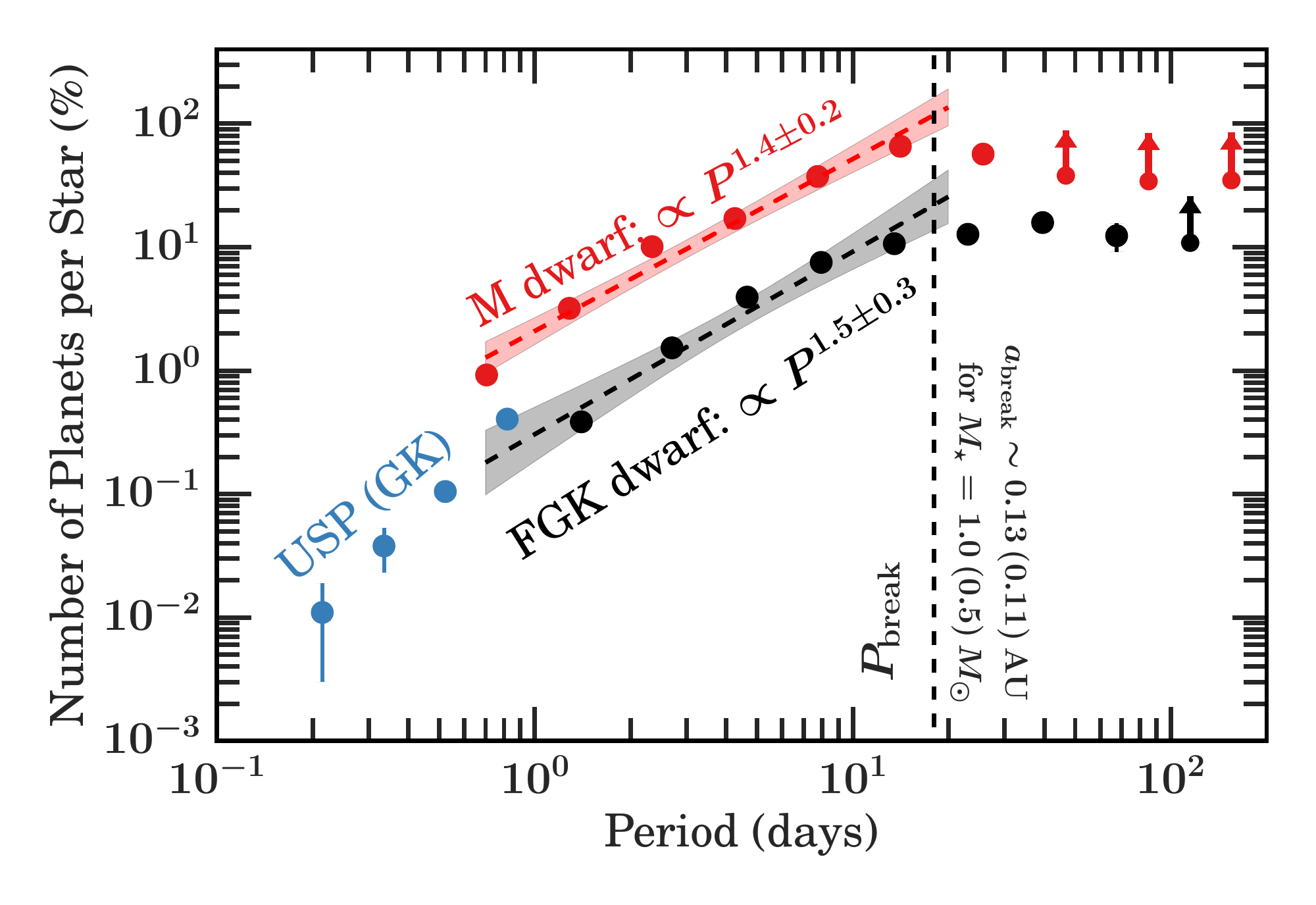}
 \end{center}
 \vspace{-0.4in}
 \caption{ From Lee \& Chiang~(2017). Occurrence of sub-Neptune
   planets as a function of period based on the {\it Kepler} studies
   of
   refs.~\cite{Fressin+2013,DressingCharbonneau2015,SanchisOjeda+2014}.
   There is no sudden change at a period of one day.  Rather, the
   period distribution changes slope at around 10 days. For M dwarfs,
   the results are plotted for bins of width $\Delta\log_{10} P =
   0.3$.  For FGK dwarfs, the bins have width 0.2 except for the point
   at $P=1.3$\,days, which has a width of 0.4.
\label{fig:evelee}}
\end{figure*}

\subsection{Period and radius distributions}
\label{subsec:pr}

Figure~\ref{fig:pr} shows the inferred radius and period distribution
of USP planets based on our FT survey, after accounting for the survey
completeness.  As we consider periods shrinking from 24 to 4 hours,
the occurrence rate falls by more than an order of magnitude.  The
trend with period is compatible with an extrapolation of the trend
that had been noted previously for periods between 1 and 10 days
\cite{Howard+2012,LeeChiang2017}. This is illustrated in
Figure~\ref{fig:evelee}.  Likewise, the occurrence rate of planets
larger than $2\,R_\oplus$ is at least a factor of five smaller than
the rate of Earth-sized planets. This, too, is compatible with a more
general trend: the radius distribution of all planets with periods
shorter than 100~days shows a dip in occurrence between 1.5 and
2\,$R_\oplus$ \cite{Fulton+2017}.

This dip has been attributed to photo-evaporation \cite{OwensWu2017}.
In this interpretation, close-in planets begin their existence as
rocky bodies of approximately $3\,M_\oplus$ which accrete differing
amounts of hydrogen and helium gas from the surrounding protoplanetary
disk.  Those that accrete only a little gas --- less than a few per
cent of the total planet mass --- lose it all during the 10$^7$ years
of high-energy irradiation by the young and magnetically active star.
Such planets are observed today as rocky bodies with sizes smaller
than $1.5\,R_\oplus$.  Most of the USP planets seem to be in this
category.  If instead a higher mass of gas is accreted, a substantial
fraction is still left over by the time the star quiets down and loses
its evaporative effect.  Such an atmosphere, even when its mass is on
the order of only a per cent of the total mass, has such a large scale
height that it increases the planet's effective size by a factor of
two.  We observe these planets today to be swollen to sizes of
2--3\,$R_\oplus$.  Such planets are commonly seen with orbital periods
from a few to 100 days, but as we have stated above, they are rarely
seen as USP planets.

\subsection{Masses}

\begin{figure*}[h!]
 \begin{center}
 \leavevmode
\includegraphics[keepaspectratio=true, width=4.5in]{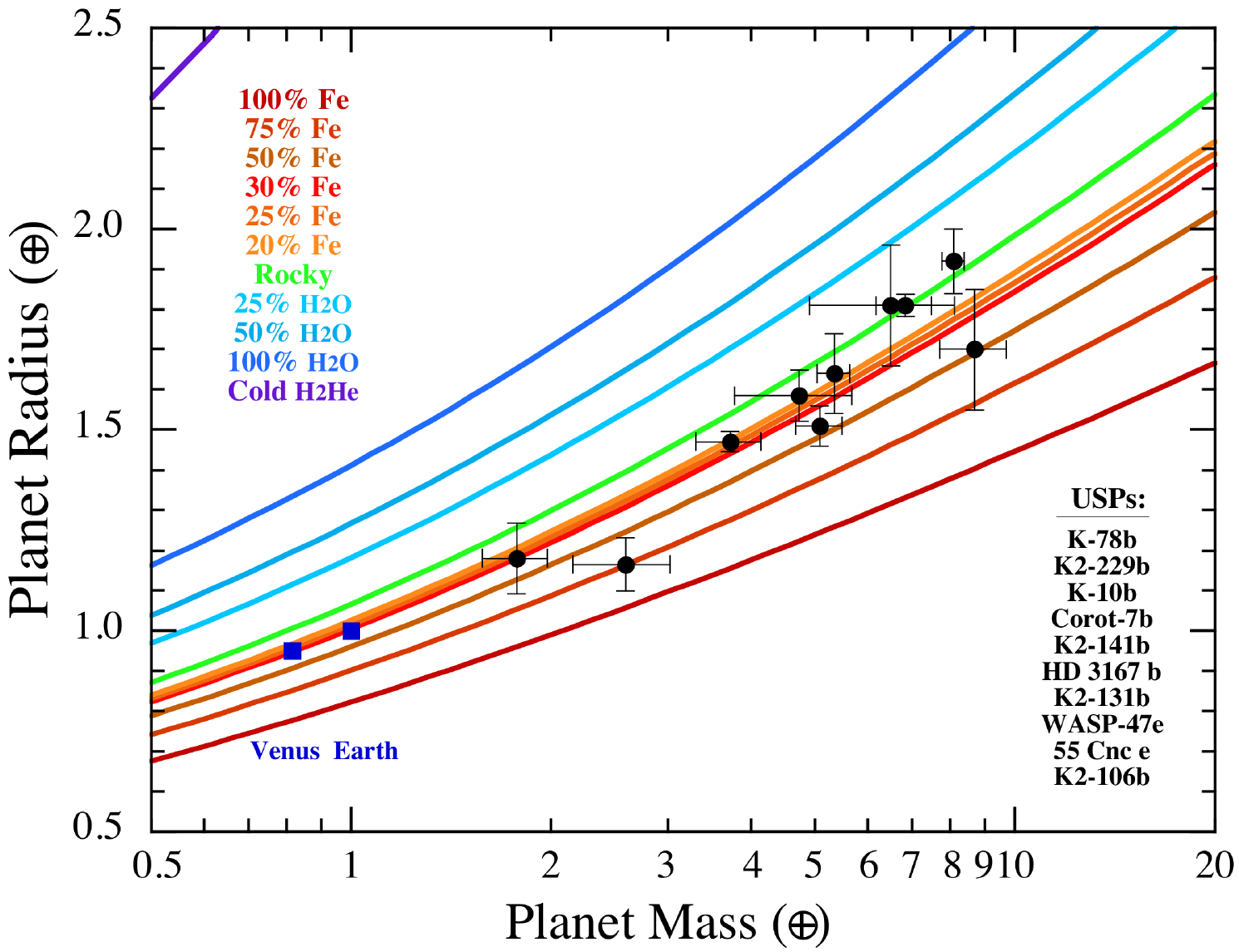}
 \end{center}
 \vspace{-0.25in}
 \caption{ Masses and radii of USP planets for which Doppler mass
   measurements have been reported. The black circles are the data.
   The planet names are printed in the lower right corner, in
   increasing order of mass.  The colored curves are theoretical
   mass/radius relationships for planets of different compositions
   \cite{Zeng+2016}: pure iron, an iron core and silicate mantle (with
   varying proportions), pure rock, silicate surrounded by a water
   envelope (with varying proportions), pure water, and cold ${\rm
     H}_2 + {\rm He}$ in cosmic proportions (the purple curve in the
   upper left corner). Here, ``rock'' and ``iron'' are shorthand for
   equations of state based on a seismically-derived model of the
   Earth's mantle and liquid core.
   References:\,\cite{
     Grunblatt+2015,     
     Weiss+2016,         
     Haywood+2014,       
     Demory+2016a,       
     Dai+2017,           
     Sinukoff+2017,      
     Guenther+2017,      
     Vanderburg+2017,    
     Gandolfi+2017,      
     Christiansen+2017,  
     Malavolta+2018,     
     Santerne+2018}.     
  \label{fig:mr}}
\end{figure*}

An important part of the Kepler-78 story was the feasibility of
Doppler mass measurement.  There are several reasons why the USP
planets are attractive targets for Doppler monitoring. The
radial-velocity amplitude scales as $P^{-1/3}$ and is therefore higher
for shorter-period planets.  A full orbit can be sampled in just a few
nights, or even a single night.  The Doppler signal is insulated from
the effects of stellar variability to some degree, because the orbital
period is usually much shorter than the stellar rotation period.
Still, because the planets tend to be small, the Doppler signals have
amplitudes of only a few meters per second, making them challenging to
detect.  Masses have been measured for ten USP planets (see
Figure~\ref{fig:mr}).  On the whole, the USP planets seem consistent
with an Earth-like composition of 70\% rock and 30\% iron.  K2-229b
has a higher density suggesting a more massive iron core.  WASP-47e
and 55\,Cnc\,e have a lower density and are compatible with pure rock,
or a rocky-iron body surrounded by a layer of water (or other
volatiles).

At the very shortest periods, a constraint on the planet's composition
can be obtained even without any Doppler data.  The mere requirement
that the planet is outside of the Roche limit --- that it has not been
ripped apart by the star's tidal gravitational force --- gives a lower
bound on the planet's mean density.  An approximate expression for
the Roche-limiting orbital period is
\begin{equation}
  P_{\rm min} \approx 12.6~{\rm hr}
  \left( \frac{\langle \rho \rangle}{1~{\rm g~cm}^{-3}} \right)^{-1/2}
  \left( \frac{\rho_{\rm c}}{\langle \rho \rangle} \right)^{-0.16},
 \label{eqn:Rochelimit}
\end{equation}
where $\langle \rho \rangle$ and $\rho_{\rm c}$ are the planet's mean and
central density, respectively.  This formula is derived from the
classical expression for the Roche limit of an incompressible fluid
body, along with a correction for compressibility (the second factor)
based on polytrope models.  It has been applied to two planets,
KOI-1843.03 ($P_{\rm orb}=4.2$~hours) and K2-137b (4.3~hours), to argue
that they probably have large iron cores
\cite{Rappaport+2013,Smith+2018}.  We note, however, that the leading
coefficient in Eqn.~(\ref{eqn:Rochelimit}) could be substantially
lower, depending on the roles of material strength and friction, and
whether the body actually splits apart once the Roche limit is
violated \cite{Davidsson1999,HolsappleMichel2006}.  More theoretical
work is warranted before we can have confidence in constraints on
planet compositions based on the Roche limit.

\subsection{Occultations}
\label{subsec:occultations}

When a planet's orbit carries it behind its host star and out of view,
the total system flux decreases by an amount proportional to the
average brightness of the planet's dayside.  As noted earlier,
Kepler-78b is the smallest planet for which it has been possible to
detect the loss of light during planetary occultations.  The next
smallest such planet is K2-141b \cite{Malavolta+2018}, which has a
size of 1.5\,$R_\oplus$ and an orbital period of 6.7~hours.  In both
cases the detection was based on white-light observations with the
{\it Kepler} telescope. In neither case has it been possible to
determine what fraction of the dayside flux arises from reflected
light, as opposed to reprocessed light (thermal emission).  This would
require data obtained over a wider range of wavelengths.

Another USP planet for which occultations have been detected is
55\,Cnc\,e (1.9\,$R_\oplus$, $P=17.7$\,hours).  In this case, the
detections were made with the {\it Spitzer} space telescope at a
wavelength of 4.5\,$\mu$m, far enough into the infrared range of the
spectrum that the signal is probably dominated by thermal emission
\cite{Demory+2012}.  The measurement was repeated eight times, and the
results for the brightness temperature ranged from 1300--2800\,K
\cite{Demory+2016a}.  The minimum and maximum amplitudes of the
observed occultation signal were found to differ by 3.3-$\sigma$.  If
this apparent variability represents genuine fluctuations of the
brightness of the planet's dayside, the observers suggested they could
be caused by widespread volcanic activity.  The observed variations in
flux over the course of the entire orbit, when attributed to the
changing planetary phase, imply dayside and nightside temperatures of
$2700\pm 270$\,K and $1380\pm 400$\,K, respectively \cite{Demory+2016b}.

\subsection{Metallicity}

Giant planets with periods shorter than a few years, including hot
Jupiters, have long been known to be more common around stars of high
metallicity. A recent {\it Kepler} study concluded that the occurrence
of hot Jupiters rises with the 3rd or 4th power of the metal abundance
\cite{Petigura+2018}.  The USP planets are also associated with
higher-than-average metallicity, but the dependence is not as strong
as for the hot Jupiters \citep{Winn+2017}.  In this respect, the USP
planets are similar to the broader sample of {\it Kepler} planets with
periods shorter than 10 days \cite{Mulders+2016, Petigura+2018,
  Wilson+2018}.

The lack of strong association between high metallicity and the
occurrence of USP planets was of particular interest because it had
been postulated that hot Jupiters are the progenitors of USP planets
\cite{Jackson+2013,Valsecchi+2015,Konigl+2017}. In this scenario, the
USP planets are the bare rocky cores of giant planets that approached
the star too closely and lost their gas, due to photo-evaporation,
Roche lobe overflow, or some other process.  However, in this
scenario, one would expect the host stars of hot Jupiters and USP
planets to have similar characteristics, including metallicity.  Since
this is not the case \cite{Winn+2017} it seems unlikely that hot
Jupiters are the progenitors of USP planets.  This still leaves open
the possibility that the progenitors are gas-ensheathed planets of
only a few Earth masses, as discussed in Section~\ref{subsec:pr}.

\begin{figure*}[h!]
 \begin{center}
 \leavevmode
\includegraphics[keepaspectratio=true, width=4.75in]{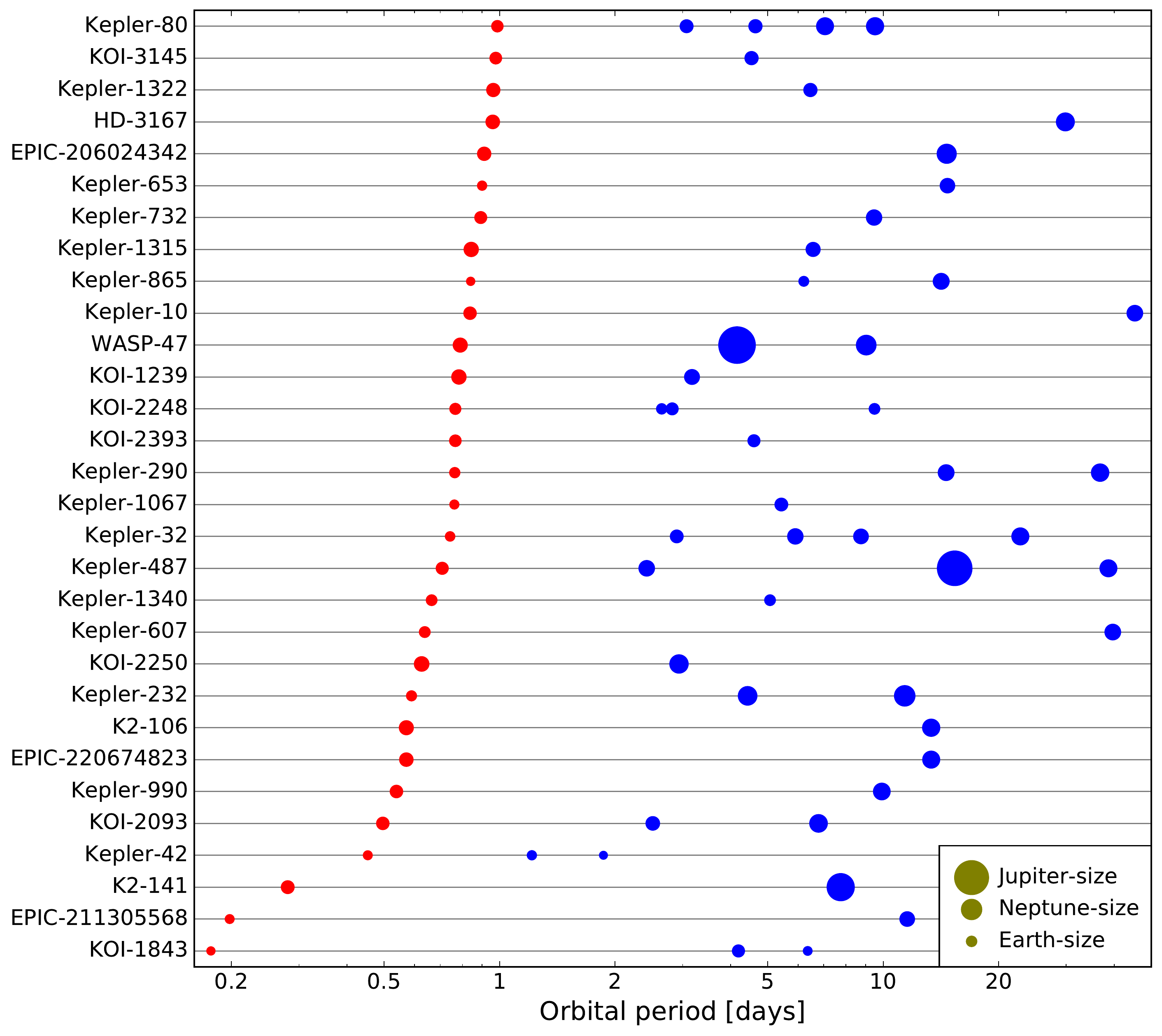}
 \end{center}
 \vspace{-0.2in}
 \caption{ Known multi-planet systems that include an
   ultra-short-period planet. Note that the period ratio between the
   innermost planet and the second planet tends to be higher than the
   other period ratios between neighboring planets. References:
   \cite{Rappaport+2013,SanchisOjeda+2014,Morton+2016,Rowe+2014,Weiss+2016,
     Twicken+2016, Malavolta+2018,Barragan+2018,Muirhead+2012,Sinukoff+2017,Adams+2017,Guenther+2017,Fabrycky+2012,Dressing+2017,Mayo+2018,Vanderburg+2016,Christiansen+2017,Becker+2015,MacDonald+2016}.
     \label{fig:usp_systems}}
\end{figure*}

\subsection{Longer period companions}

Another difference between hot Jupiters and USP planets is in
their tendency to have nearby planetary companions.  Hot Jupiters are
rarely found with other planets within a factor of 2--3 in orbital
period or distance \cite{Steffen+2012}.  In contrast, USP planets are
almost always associated with longer-period planetary companions
\cite{SanchisOjeda+2014,Adams+2017}.  This conclusion is partly based
on the numerous detections of longer-period transiting planets in
systems that have a USP planet (see Figure~\ref{fig:usp_systems}).  It
is also based on a statistical argument involving the decrease in
transit probability with orbital period: many of the {\it Kepler}
stars for which only the transits of a USP planet are detected must be
multi-planet systems for which the outer planet does not happen to be
transiting \cite{SanchisOjeda+2014}.

One exceptional system is WASP-47, which has a hot Jupiter and a USP
planet \cite{Becker+2015}.  There is also a third planet orbiting just
outside of the orbit of the hot Jupiter.  This system seems too
dynamically fragile for the hot Jupiter to have undergone
high-eccentricity migration, a scenario that is often invoked to
explain how a giant planet could find its way into such a tight orbit.
A similar system is Kepler-487, which has a USP planet and a ``warm
Jupiter'' with a period of 15.4~days, in addition to two other
transiting planets \cite{Rowe+2014}. This system has not yet been as
well characterized as WASP-47.

The multi-planet systems that include a USP planet differ from the
other {\it Kepler} multi-planet systems in an intriguing way.  A
typical system without a USP planet has a ratio of between 1.5 and 4
between the periods of the outer and inner planets of an adjacent pair
\cite{Fabrycky+2014}.  But when the inner planet has a period shorter
than one day, the period ratio is almost always greater than 3
\cite{Steffen+2013}, as illustrated in Figure~\ref{fig:usp_systems}.
This suggests that some process has widened the period ratio.  Perhaps
the period of the USP planet has shrunk due to tidal orbital decay
that is either ongoing, or that took place early in the system's
history when the star had still not contracted onto the main sequence
and would have been more susceptible to tides.

The same systems that show wider-than-usual period ratios also appear
to have higher mutual inclinations. This is based on comparisons
between the transit impact parameters of planets in the same system,
which gives a lower limit on mutual inclination.  Among the {\it
  Kepler} multi-planet systems for which the innermost planet has
$a/R_\star<5$, the mutual inclination distribution ranges up to 10-15
degrees \cite{Rodriguez+2018,Dai+2018}. This is higher than the 2-5
degrees that have generally been found for pairs of planets in wider
orbits \cite{Fabrycky+2014}. This finding suggests that USP planets
have experienced inclination excitation in addition to orbital
shrinkage.

\subsection{Formation theories}

Before describing some of the theories for the formation of USP
planets, it is time for another reminder that there is no sharp
astrophysical distinction between the ``ultra-short-period'' planets
with periods of less than one day and the merely ``short-period''
planets with periods of 1--10 days.  The problems related to the
formation of short-period planets have been with us since 1995, and we
will not attempt a comprehensive review here.  Instead we make note of
some of the work relating specifically to the planets with the very
shortest orbital periods.

Even before the discovery of Corot-7b, Raymond et al.\ enumerated six
possible formation pathways for ``hot Earths'' \cite{Raymond+2008}.
As summarized below, these pathways lead to different predictions for
the final compositions for the USP planets, and for the properties of
any additional planets around the same star.  In light of the
knowledge we have gained over the last 10 years, we can try to fill in
the score card:
\begin{enumerate}
\item Accretion from solid material in the innermost part of the
  protoplanetary disk. This ``{\it in situ} accretion'' process would
  typically lead to several hot Earths spaced by 20--60 mutual Hill
  radii.  Such systems have indeed been observed.  This formation
  pathway would also result in a dry composition (0.1--1\% water),
  which is consistent with the existing mass and radius measurements.

\item Spiraling-in of a planet from beyond the ice line, due to
  gravitational interactions with the disk (Type I migration).  This
  would likely lead to resonant chains of multiple planets.  Such
  resonant chains are infrequent in the {\it Kepler} sample, and to
  our knowledge there is are no examples involving a USP planet.  This
  scenario would also lead to water-rich planets, with 10\% or more of
  their mass in water.  The observations are not yet good enough to
  try and confirm or rule out a water fraction of 10\%.

\item Accretion of material that is locked in mean-motion resonance
  with a migrating giant planet and thereby ``shepherded'' inward.
  This would lead to systems in which the USP planet is parked next to
  a giant planet.  The only known systems that fit this description are
  WASP-47 and Kepler-487.

\item Accretion of material shepherded by sweeping secular resonances.
  In this scenario, there are at least two giant planets. The
  resonance is between the precession rate induced by the other
  planet, and that induced by the (gradually disappearing) disk. This
  would lead to systems in which hot Earths and two giant planets
  coexist, which have not been observed.

\item Eccentricity excitation followed by tidal circularization of an
  initially wide-orbiting planet.  This would result in isolated USP
  planets.  In reality, there are many USP planets with known
  companions (see Fig.~\ref{fig:usp_systems}), and statistical arguments suggest
  that the majority of USPs are part of compact multi-planet systems
  \cite{SanchisOjeda+2013,Adams+2017}.
  
\item Photo-evaporation of a formerly gaseous planet that approached
  the star too closely.  As noted earlier, this scenario was later
  elaborated to predict a gap in the radius distribution of close-in
  planets --- or a ``valley'' in the space of radius and period ---
  which has been observed.  Of course, this theory by itself does not
  explain where the progenitor planet came from.
\end{enumerate}

Obviously the picture is not yet clear, nor is it clear that these six
pathways are the only possibilities.  Nevertheless, it does seem that
elements of the {\it in situ} and photo-evaporation theories have
withstood a decade of observations.  In recent years, a few more
specific theories have been offered:

\begin{enumerate}
  
\item Schlaufman et al.\ proposed that a planet can be driven to short
  periods by dynamical interactions with nearby planets in wider
  orbits, at which point tidal interactions with the stars shrink the
  orbit still further, forming an ultra-short-period planet
  \cite{Schlaufman2010b}.  They predicted that USP planets should be
  more common around massive (F-type) stars than around lower-mass
  stars, because massive stars should have weaker tidal dissipation.
  This appears to be the opposite of what is observed; in our survey,
  the occurrence was lowest for the F dwarfs and highest for M dwarfs.

\item Lee \& Chiang agreed that tidal dissipation is responsible for
  shrinking the orbits, but proposed a different initial condition:
  the planet forms from material that collects near the innermost edge
  of the protoplanetary disk \cite{LeeChiang2017}.  The location of
  the inner edge is the distance at which the orbital period matched
  the stellar rotation period at early times, which is 10~days for
  Sun-like stars. In this way, they explain why the occurrence rate of
  planets begins dropping for periods less than 10 days.  In their
  theory, the USP planets are gradually spiraling inward due to tidal
  orbital decay.  They predict that for more massive and rapidly
  rotating A stars, the break in the period distribution of close-in
  rocky planets (should they exist) will occur closer to one day.

\item Petrovich et al.\ investigated the possibility of eccentricity
  excitation from secular dynamical chaos in compact multiplanet
  systems \cite{Petrovich+2019}.  They predicted that USP planets
  would often be accompanied by outer planets extending out to
  1~AU. They also predicted that USP planets would show a broad range
  of inclinations with respect to the equator of the host star, and
  mutual inclinations with respect to the orbits of outer
  planets. This latter prediction, at least, has found empirical
  support \cite{Dai+2018}.

\end{enumerate}

\section{Other ultra-short-period phenomena}
\label{sec:other}

\subsection{Giant planets}

There are 6 known examples of giant planets ($R_{\rm p} >
8\,R_\oplus$) with a period shorter than one day. These rare
butterflies are WASP-18b, WASP-43b, WASP-103b, HATS-18b, KELT-16b, and
WASP-19b, the last of which has the shortest known period of 0.788
days \cite{Hebb+2010}. None of them were found in the {\it Kepler}
survey; rather, they were found in ground-based transit surveys that
were capable of searching a larger sample of stars. Their occurrence
must be lower by at least an order of magnitude than that of
terrestrial-sized USP planets.  Because of the large masses of these
planets, they are valuable for testing theories of gravitational tidal
interactions.  As described above and in Section~\ref{subsec:basic},
tides should cause the star to spin faster and the orbit to shrink,
ultimately leading to the engulfment of the planet.  However, tidal
theory makes no firm prediction for the timescale over which the orbit
decays. The timescale depends not only on the planet mass and orbital
distance, but also on the rate of tidal dissipation within the star,
which is uncertain by at least an order of magnitude.  If we could
observe the steady shrinkage of the orbital period of a hot Jupiter,
we would be able to confirm a fundamental theoretical prediction and
clarify the rate of tidal dissipation within Sun-like stars, a
longstanding uncertainty in stellar astrophysics. So far, the best
candidate for period decay is WASP-12
\cite{Maciejewski+2016,Patra+2017}.


\subsection{Pseudoplanets}

The T Tauri star PTFO 8-8695, less than a few million years old,
exhibits periodic fading events that were interpreted as the transits
of a giant planet on a precessing orbit \cite{VanEyken+2012,
  Barnes+2013,Ciardi+2015}. This discovery was greeted with great
interest, because the study of hot Jupiters around very young stars
would provide information about the timing of planet formation, the
structure of newborn planets still cooling and contracting, and the
mechanism for shrinking planetary orbits and creating hot Jupiters.
However, follow-up observations revealed some problems with the planet
hypothesis: the ``transits'' are not strictly periodic, the shape of
the light curve varies substantially with wavelength and orbital
cycle, and an occultation signal with the expected amplitude has been
ruled out \cite{Yu+2015}. The origin of the fading events is still
unknown, though, and some investigators are still pursuing the planet
hypothesis \cite{JohnsKrull+2017}.

Another case of questionable USP planets is KIC\,05807616, a hot B
subdwarf that showed evidence for two planets with orbital periods 5.8
and 8.2~hours.  The evidence was based on the observed periodic
modulations in the light from the system, which were consistent with
the illumination variations of the putative planets
\cite{Charpinet+2011}. However, additional data showed that these
periodic variations were not coherent enough to arise from orbital
motion \cite{Krzesinski2015}.  Again, the true origin of these
planet-like signals has not been ascertained.

\subsection{Disintegrating planets}

Another discovery that emerged from the Fourier-based search of the
{\it Kepler} data was the first of a small subset of objects referred
to as ``disintegrating" planets.  The objects are KIC\,12557548b
(KIC\,1255b for short), KOI-2700b, and K2-22b, with orbital periods of
15.7, 22, and 9.5 hours, respectively \cite{Rappaport2012a,
  Rappaport2014a, SanchisOjeda+2015b, vanLieshout2017}.  In all three
cases, the transit signal is asymmetric around the time of minimum
light, and there are variations in the transit depth.  KIC\,1255b and
KOI-2700b have much longer egress times than ingress times, while
KIC\,1255b and K2-22b exhibit highly variable depths, often from one
transit to the next.

These characteristics strongly suggest that the occulter is an
elongated tail of dusty material streaming away from a hot rocky
planet. It is important to note that the transits in these objects do
not reveal any evidence of the underlying solid body itself.  For two
of the objects the transit depths are on the order of 0.5\%, implying
an obscuring area comparable to that of Jupiter, but the obscuration
is presumed to be almost entirely due to dust extinction. In all three
cases the upper limits on the size of the solid body itself is on the
order of the size of the Earth, and theoretical considerations suggest
the true sizes may be comparable to that of the Moon
\cite{PerezBeckerChiang2013}.

The shape of the dust tail is largely dictated by radiation pressure
forces.  The likely result is a trailing dust tail unless the
radiation pressure forces are small, which can occur for either very
large ($\gsim$\,10\,$\mu$m) or tiny dust particles
($\lsim$\,0.1\,$\mu$m). The inferred mass-loss rates from the planets
are based on the amount of dust required to yield such significant
extinction of the host star, and are in the range of
$10^{10}$--$10^{11}$~g~s$^{-1}$. This is roughly equivalent to a few
lunar masses per Gyr.  These and other properties have been recently
reviewed by Van~Lieshout \& Rappaport \cite{vanLieshout2017}.

\subsection{Disintegrating asteroids}

WD\,1145+017 is a unique object thought to be a white dwarf with a set
of disintegrating asteroids in ultra-short-period orbits
\cite{Vanderburg+2015}.  The system exhibits transits with multiple
periods in the range from 4.5 to 4.9 hours.  It is believed that the
asteroids are responsible for dust clouds which then produce the
transits.  These transits can be as deep as 60\% and last anywhere
from about 10 minutes to an hour. All the bodies that have been
inferred are orbiting the white dwarf at a distance of only one
stellar radius. Given the stellar luminosity of 0.1\,$L_\odot$, the
equilibrium temperatures of the dust grains are about 1000--2000\,K,
depending on their size and composition \cite{Xu+2018}. We are likely
witnessing the disintegration of planetary bodies that survived the
metamorphosis of the host star from a dwarf into a giant and then
into a white dwarf.

\subsection{Unknown unknowns}

The disintegrating planets and asteroids are examples of new and
interesting phenomena that were discovered by sifting through {\it
  Kepler} data.  The next few years will bring another good
opportunity for a comprehensive exploration of ultra-short-period
phenomena.  The {\it Transiting Exoplanet Survey Satellite} ({\it
  TESS}\,), launched in April 2018, is performing time-series
photometry over about 90\% of the sky using four 10~cm telescopes
\cite{Ricker+2015}.  Stars are observed for a duration ranging
from one month to one year, depending on ecliptic latitude.

With {\it TESS} data, we will be able to find USP planets around stars
that are several magnitudes brighter than typical {\it Kepler}
stars. This will provide more targets that are suitable for precise
Doppler mass measurements.  With a sample of brighter stars, it will
also be easier to test for compositional similarities between stars
with USP planets and stars with other types of planets that might be
the progenitors of USP planets.

More generally, the {\it TESS} data will be another giant haystack
within which to search for interesting needles: planets in unusual
configurations, disintegrating planets and asteroids, and hitherto
unknown phenomena.  Searches for USP planets may reveal the
hypothesized ``iron planets'' that could be formed when rocky planets
are battered by giant impacts, vaporizing the rocky mantle but leaving
the iron core intact.  {\it TESS} will also be able to search for hot
Jupiters within a sample of stars that is orders of magnitude larger
than the {\it Kepler} sample, perhaps allowing us to find planets
undergoing rapid orbital decay.  Although {\it TESS} attracts the most
attention for its potential to find potentially habitable planets
around red dwarf stars, the mission can also be regarded as a search
for rare short-period phenomena over a wider range of stellar masses
and ages than has been explored before.

\vspace{0.5in} {\it Acknowledgements.}---We thank Jack Lissauer for
the invitation to write this review, and Scott Tremaine for helpful
consultations.  We thank Eve Lee and Eugene Chiang for granting us
permission to reproduce Figure~1 from their 2017 paper (which appears
as Figure~\ref{fig:evelee} in this article), and Jerry Chen for
pointing out an error in Eqn.~(4) in the originally published version
of this article. J.N.W.\ thanks the Heising-Simons foundation for
support. We are all deeply grateful to the engineers, managers, and
scientists who were reponsible for the {\it Kepler} mission which has
led to so many important discoveries.

\section*{References}

\bibliography{references}

\end{document}